# Twinning superlattices in indium phosphide nanowires

R.E. Algra[1,2,3], M.A. Verheijen[2], M.T. Borgström[2,4], L.F. Feiner[2], G. Immink[2], W.J.P. van Enckevort[3], E. Vlieg[3], E.P.A.M. Bakkers[2,*]

[1]Materials Innovation Institute (M2i), 2628CD Delft, The Netherlands

[2]Philips Research Laboratories Eindhoven, High Tech Campus 11, 5656AE Eindhoven, The Netherlands

[3]IMM, Solid State Chemistry, Radboud University Nijmegen, Heijendaalseweg 135, 6525AJ Nijmegen, The Netherlands

[4] Solid State Physics, Lund University, Box 118, S-221 00 Lund, Sweden

*E-mail: erik.bakkers@philips.com

**Semiconducting nanowires offer the possibility of nearly unlimited complex bottom-up design [1,2], which allows for new device concepts [3,4]. However, essential parameters that determine the electronic quality of the wires, and which have not been controlled yet for the III-V compound semiconductors, are the stacking fault density [5] and the wire crystal structure. In addition, a significant feature would be to have a constant spacing between rotational twins in the wires such that a twinning superlattice (TSL) is formed, since this is predicted to induce a direct bandgap in normally indirect bandgap semiconductors [6,7], such as silicon and gallium phosphide. Optically active versions of these technologically relevant semiconductors will have major impact on the electronics [8] and optics [9] industry. Here, we show that we control the crystal structure of indium phosphide (InP) nanowires by impurity dopants. We have found that zinc decreases the activation barrier for 2D nucleation growth of**

**zinc-blende InP and therefore promotes the InP nanowires to crystallise in the zinc blende, instead of the commonly found wurtzite crystal structure [10]. More importantly, we demonstrate that we can, by controlling the crystal structure, induce twinning superlattices with long-range order in InP nanowires. We can tune the spacing of the superlattices by the wire diameter and the zinc concentration and present a model based on the cross-sectional shape of the zinc-blende InP nanowires to quantitatively explain the formation of the periodic twinning.**

Twin planes and, more generally, planar stacking faults are commonly found in III-V nanowires grown in the [111] direction by the vapour-liquid-solid (VLS) mechanism. A twin plane in a zinc-blende (ZB) (stacking fault in wurtzite (Wz)) nanowire can be considered as a monolayer of the Wz (ZB) phase [11]. Stacking faults can significantly affect the electronic properties of the nanowires [5,7]. The electron wavefunction is discontinuous at a stacking fault which leads, for instance, to a reduced mobility of charge carriers. The formation and resulting morphology of randomly distributed stacking faults in nanowires have been investigated by several authors [11-14]. Twin planes that have a constant spacing within a nanowire form a twinning superlattice (TSL); this modifies the electronic band structure, giving rise to the formation of minibands [7]. Recently, small domain TSLs have been observed locally in bulk Si [15,16], and occasionally in ZnS nanowires [17], but the parameters controlling the phenomenon were not identified.

We have synthesized InP nanowires from colloidal gold particles by VLS growth using metal-organic vapour phase epitaxy (MOVPE) with trimethylindium and phosphine as molecular precursors, as described in the Methods Section. To establish p-type doping in our nanowires we introduce diethylzinc (DEZn) in the growth system. For Zn partial pressures below $4.6*10^{-4}$ mbar ($4.6*10^{-4}$ mbar corresponds to a free hole concentration of $10^{18}$ cm$^{-3}$ in the InP nanowires [18]) we find that twin planes are randomly distributed in the nanowires. Strikingly, above $4.6*10^{-4}$ mbar the twin planes exhibit a constant spacing for a given Zn concentration and wire diameter, and the nanowire develops into a periodically twinned superlattice. The segment length of the periodic structure increases with Zn concentration and wire diameter. In Figure 1, transmission electron micrographs (TEM) are shown of Zn-doped ($9.2*10^{-4}$ mbar) InP nanowires with nominal diameters of 10, 20, 50 and 100 nm. It is clear from the overview images in Figure 1A that the periodically twinned structure is general, although not all of the wires have the optimal orientation with respect to the electron beam. The segment length is uniform (Figure 1B) throughout the wires almost up to the catalyst particle. From the high resolution images in Figure 1C we observe that the periodic nanowires have the ZB crystal structure with non-parallel {111} side facets with respect to the long nanowire axis (see Figure 1C). From the high resolution images the number of monolayers between successive twin planes is counted and the data is plotted in the histograms shown in Figure 2A (here, a monolayer with thickness $d_{111}$=3.4 Å contains pairs of In and P atoms). Segment lengths of 7±3, 13±3, 26±4 and 35±3 monolayers were found for wires with a diameter of 10, 20, 50 and 100 nm, respectively, see Figure 2B. Importantly, the periodicity in twinning is demonstrated by the relatively narrow distributions in segment lengths.

These results demonstrate that the wire diameter and the Zn doping are important parameters controlling the periodicity. We will now first discuss the effect of Zn on the crystal structure and then present a model based on the evolution of the cross-sectional wire shape to quantitatively explain the formation of the periodic structures.

Bulk InP has the ZB crystal structure, because the free energy is slightly lower ($\Delta E$=6.8 meV/III-V atom pair) [19] for ZB than for Wz InP. However, nominally undoped InP nanowires commonly exhibit the wurtzite crystal structure. Possible explanations for the formation of Wz nanowires are the lower surface energy of the parallel side facets of Wz wires compared to that of ZB wires [19] and the interface energies at the V-L-S three-phase line [20]. These effects would make crystallisation in the Wz phase especially favourable for thin wires that have a large surface to bulk ratio. However, we consistently observe that undoped InP wires (with diameters from 10 up to 250 nm) have Wz structure, though in general contain stacking faults. With increasing diameter the number of stacking faults decreases, leading to wires with a larger fraction of Wz and showing that the above mentioned factors do not ultimately determine the crystal structure of the wires.

The main difference between bulk and VLS growth is the presence of the catalyst particle from which the crystal is precipitated, and therefore the atomic interactions at the liquid–solid interface should be considered. We find that the parameter critically determining the nanowire crystal structure and the stacking fault density is the chemical composition of the catalyst particle near the liquid-solid interface. The high number of planar stacking faults in the undoped InP Wz nanowires can be reduced by adding sulphur (S) to the gas phase. At the highest S partial pressure of $4.2*10^{-3}$ mbar ($8.3*10^{-7}$ mbar corresponds to a free electron concentration of $3 \cdot 10^{18}$ cm$^{-3}$ in our nanowires) [18] a perfect Wz crystal without stacking faults was obtained (see supplementary information S1, Figure S1). In contrast, when sufficient Zn is added to the system, the nanowires precipitate in the ZB crystal structure. We find a transition from Wz to ZB at a DEZn partial pressure of $4.6*10^{-5}$ mbar as is shown in Figure 2C.

In order to quantify the effect of Zn on the crystal structure we have calculated, based on 2D nucleation, a kinetic phase diagram, as shown in Figure 2D, separating the domains of Wz and ZB nanowire growth with respect to the supersaturation in the droplet, $\Delta\mu$, and the (normalized) difference in solid-liquid step free energy between a ZB and a Wz nucleus, $\Delta\gamma/\gamma_{sl,ZB}$ (see supplementary information S2). As elaborated in the supplementary information, the main effect of adding zinc during growth is a decrease in $\Delta\gamma/\gamma_{sl,ZB}$, which means a lowering of the liquid-solid step energy for ZB as compared to Wz. This suggests a strong interaction of the zinc atoms with the InP growth interface as was also observed during electrical resistance measurements of Au(Zn)-InP contacts[21,22]. In general, for nanowires grown by the VLS mechanism the crystal structure may be intrinsically different from that of the bulk material and will depend on the combination of semiconductor with catalyst material.

The Zn-doped InP nanowires have the ZB crystal structure with non-parallel {111} side facets, tilted by $\theta = \theta_B = -\theta_A \approx 19.5°$ with respect to the long nanowire axis (see Figure 1C), which is crucial for the formation of the twinning superlattice. As shown in Figure 3, at a certain moment during growth, 1, the top surface of the nanowire is a hexagon, and the catalyst droplet is only slightly deformed from a spherical shape. When the wire grows, the {111}A edges move inward and their length increases, while the {111}B edges move outward and their length decreases. Thus the shape of the nanowire-droplet interface becomes increasingly triangle-like, as shown for situation 2 in Figure 3. As a consequence, the catalyst droplet distorts, and at a certain point it becomes energetically more favourable to form a twin plane and to start reducing the distortion of the catalyst particle by re-growth towards a hexagonal shape, rather than to continue growth towards a completely triangular shape. This mechanism of inverting triangularly shaped interfaces repeats itself continuously and produces the periodically structured wire.

We have analyzed this behaviour quantitatively, making use of the simulator program 'Surface Evolver' [23,24] to calculate the distortion of the droplet. Crucial observations are (see Figure 3B), (i) that the contact angle between the droplet and the top surface of the nanowire at the {111}A edges becomes different from the contact angle at the {111}B edges, and (ii) that their difference depends linearly on the deformation of the droplet-nanowire interface and therefore also on the ratio $H/D$ of the wire height $H$ (measured from the last hexagonal cross section) to the wire diameter $D$ (see supplementary information S3). Since nucleation is initiated at the nanowire edge [12], the contact angles affect the 2D nucleation processes involved in the nanowire growth. Because the free energy of formation is significantly lower for nuclei with an external (i.e. solid-vapour) B-facet than for those with an external A-facet [25] (see also supplementary information S3 and Figure S5), the probability of formation of B-facet nuclei is strongly enhanced. As a result, the occurrence of twins is determined by the competition between B-facet nuclei nucleating at B edges, adding another ZB layer, and at A edges, introducing a layer that involves a twin plane and initiating re-growth. It follows that the critical height $H_c$, at which twin formation becomes the more favourable process, is proportional to the wire diameter $D$, in apparent agreement with what has been observed for Si nanowires and tentatively explained by considerations based upon total energies instead of nucleation energies [26]. However, the segment length will be substantially less than $2H_c$, because it is determined by the probability of an uninterrupted series of facet-conserving nucleations, and not by that for a single facet-changing nucleation. Taking this statistical aspect into account we find that the number of layers in a segment is given by $N_s = AD - 2 + D/B \ln[\exp(2B/D) - 1]$. We find excellent agreement between this expression and the observed diameter dependence, as shown in Figure 2B, if we make use of the explicit expressions [25] of the constants $A$ and $B$ in terms of the physical parameters (solid-liquid surface tension $\gamma_{SL}$, liquid-vapour surface tension $\gamma_{LV}$, surface energy of a twin plane $\gamma_T$, supersaturation $\Delta\mu$, tilting angle $\theta$, contact angle at hexagonal interface shape $\beta_0$, and temperature $T$), using $T = 713$ K, $\gamma_T = 0.009$ J/m$^2$ [19, 27], $\gamma_{SL} \approx \gamma_{SV} \approx 0.8$ J/m$^2$ [28], $\gamma_{LV} = 1.0$ J/m$^2$ (in between the values for liquid Au and In), with $\beta_0 \approx 98°$, and $\Delta\mu = 180$ meV/atom pair, which are physically plausible values, corresponding to a critical nucleus with a diameter of about 2.8 nm.

Our insight in the formation of twins allows the fabrication of more complex structures by varying the Zn concentration during growth. Without Zn, random twinning should occur, and with Zn present the twinning should become periodic. This is indeed the case as demonstrated by the TEM images in Figures 4A and 4B, showing a wire in which intermittently a Zn partial pressure of 9.2*10$^{-4}$ mbar has been used. The tapering of the nanowire is due to sidewall growth, which preferentially occurs on the ZB sections. In Figure 4C, the length of the ZB sections has been plotted versus the time interval during which the Zn precursor gas flow was switched on. The fitted linear curve has an insignificant offset from zero, suggesting an almost immediate switching from the Wz to the ZB phase and vice versa. In Figure 4D the number of monolayers between two twin planes is presented for the first four ZB sections. The average segment length is 13±3 monolayers, showing that the periodicity is clearly preserved in these short sections and is not affected by the growth history.

We have presented a viable route for the fabrication of twinning superlattices in nanowires by controlling the nanowire morphology. This new instrument for manipulating the electronic properties of nanowires can be combined with already demonstrated features such as axial and radial heterostructures and doping profiles, further expanding the nanowire toolbox.

**Methods**

The InP nanowires were synthesized in a low pressure (50 mbar) Aixtron 200 MOVPE reactor on InP (111)B substrates. The substrates were treated with a piranha etch for 1 min to remove the surface oxide, before deposition of Au colloids of different diameters (ranging from 10 to 200 nm). The nanowires were grown in the VLS growth mode using trimethylindium (TMIn) and phosphine ($PH_3$) as precursors, $1.19*10^{-3}$ and $4.17*10^{-1}$ mbar, respectively, in a total flow of 6 L min$^{-1}$ hydrogen ($H_2$) carrier gas. As dopant materials diethylzinc (DEZn) and dihydrogensulfide ($H_2S$) were used for p-type and n-type doping, respectively. Before growth an anneal step was carried out under $PH_3/H_2$ atmosphere to desorb any surface oxide and alloy the Au colloids with the InP substrate to ensure epitaxial growth. Growth was initiated when a temperature of 420°C was reached by switching on the TMIn. To change the dopant concentrations in the wires the DEZn and $H_2S$ partial pressures were varied between $10^{-2}$ and $10^{-7}$ mbar, at constant TMIn and $PH_3$ molar fractions in $H_2$. We like to point out that the used molar fractions for DEZn and $H_2S$ are controlled gas flows in the reactor and not necessarily the built-in atomic fractions in the wires. The samples were analyzed using transmission electron microscopy (TEM FEI Tecnai 300 kV) in bright field, and high resolution (HRTEM).

**Acknowledgements**

This research was carried out under project number MC3.05243 in the framework of the strategic research program of the Materials Innovation Institute (M2i) ([www.M2i.nl](www.M2i.nl)), the former Netherlands Institute of Metals Research, the FP6 NODE (015783) project, the ministry of economic affairs in the Netherlands (NanoNed) and the European Marie Curie program. The authors thank H. de Barse and F. Holthuysen for SEM imaging and Paul van der Sluis and Harry Wondergem for useful discussions. Correspondence and requests for materials should be addressed to E.P.A.M. Bakkers.

**Author contributions**

All authors contributed to the design of experiments. G.I. is responsible for MOVPE growth, and M.A.V. for the TEM experiments. R.E.A. and M.A.V. analysed the TEM data. L.F.F. and W.J.P.v.E. analysed the data quantitatively. R.E.A., L.F.F., W.J.P.v.E. and E.P.A.M.B. co-wrote the paper.

**References**

1. Gudikson, M.S. et al. Growth of nanowire superlattice structures for nanoscale photonics and electronics. *Nature* **415**, 617-620 (2002)
2. Dick, K.A. et al. Synthesis of branched 'nanotrees' by controlled seeding of multiple branching events, *Nature Mat.* **3**, 380-384 (2004)

3. Hochbaum, A.I., Enhanced thermoelectric performance of rough silicon nanowires, *Nature* **451**, 163-168 (2008)

4. van Dam, J.A., Supercurrent reversal in quantum dots, *Nature* **442**, 667-670 (2006)

5. Bao, J. et al. Optical properties of rotationally twinned InP nanowire heterostructures, *Nano Letters* **8**, 836-841 (2008)

6. Ikonic, Z. et al. Electronic properties of twin boundaries and twinning superlattices in diamond-type and zinc-blende-type semiconductors. *Phys. Rev. B* **48**, 17181-17193 (1993)

7. Ikonic, Z. et al. Optical properties of twinning superlattices in diamond-type and zinc-blende-type semiconductors. *Phys. Rev. B* **52**, 14078-14085 (1995)

8. Lui, A.S. et al. Advances in silicon photonic devices for silicon based optoelectronic applications. *Physica E* **35**, 223-228 (2006)

9. Krames, M.R. et al. Status and future of high-power- light-emitting diodes for solid-state lighting. *J. Display Tech.* **3**, 160-175 (2007)

10. Mattila, M. et al. Crystal-structure-dependent photoluminescence from InP nanowires. *Nanotechnology* **17**, 1580-1583 (2006)

11. Hiruma, K. et al. Growth and optical properties of nanometer scale GaAs and InAs whiskers. *J. Appl. Phys.* **77**, 447-462 (1995)

12. Johansson, J. et al. Structural properties of <111>B-oriented III-V nanowires. *Nature Mat.* **5**, 574-580 (2006)

13. Xiong, Q. et al. Coherent twinning phenomena: Towards twinning superlattices in III-V semiconducting nanowires. *Nano Letters* **6**, 2736-2742 (2006)

14. Verheijen, M.A. et al. Three dimensional morphology of GaP-GaAs nanowires revealed by transmission electron microscopy tomography. *Nano Letters* **7**, 3051-3055 (2007).

15. Fissel, A. et al. Formation of twinning-superlattice regions by artificial stacking of Si layers. *J. Cryst. Growth* **290**, 392-397 (2006)

16. Hibino, H. et al. Twinned epitaxial layers formed on Si(111) √3x√3-B. *J. Vac. Sci. Technol. A* **16**, 1934-1937 (1998)

17. Hao, Y. et al. Periodically twinned nanowires and polytypic nanobelts of ZnS: The role of mass diffusion in vapor-liquid-solid growth. *Nano Letters* **6**, 1650-1655 (2006)

18. Minot, E.D. et al. Single quantum dot nanowire LEDs. *Nano Letters* **7**, 367-371 (2007)

19. Akiyama, T. et al. An empirical potential approach to wurtzite- zinc-blende polytypism in group III-V semiconducting nanowires. *J. J. Appl. Phys.* **45**, L275-L278 (2006)

20. Glas, F. et al. Why does Wurtzite form in nanowires of III-V Zinc Blende semiconductors? *Phys. Rev. Lett.* **99**, 146101 (2007)

21. Malina, V. et al. Effect of deposition parameters in the electrical and metallurgical properties of Au-Zn contacts to p-type InP. *Semicond. Sci. Technol.* **9**, 1523-1528 (1994)

22. Weizer, G.W. et al. Au/Zn contacts to ρ-InP: Electrical and metallurgical characteristics and the relationship between them. *NASA Technical Memorandum* 106590 (1994)

23. Brakke, K.E. The surface evolver. *Experimental Mathematics* 1, 141-165 (1992)

24. http://www.susqu.edu/brakke/evolver/evolver.html

25. Feiner, L.F.  To be published.

26. Ross, F.M. et al. Sawtooth faceting in silicon nanowires. *Phys. Rev. Lett.* **95**, 146104 (2005)

27. Hurle, D.T.J.  A mechanism for twin formation during Czochralski and encapsulated vertical Bridgman growth of III-V compound semiconductors. *J. Cryst. Growth* **147**, 239-250 (1995)

28. Liu, Q.K.K. et al. Equilibrium shapes and energies of coherent strained InP islands. *Phys. Rev. B* **60**, 17008 (1999)

# Figure Captions

**Figure 1: Transmission electron micrographs of nanowire twinning superlattices.** A, B) Overview and C) high resolution TEM images of InP nanowires with a diameter of nominally 10, 20, 50, and 100nm and a DEZn partial pressure of $9.7*10^{-4}$ mbar. The scalebars correspond to A) 100 nm, B) 50 nm, and C) 5 nm.

**Figure 2: The effect of wire diameter and zinc doping on the twin lattice spacing.** A) Histograms of the number of monolayers between two consecutive twin planes for wires with a diameter of 10, 20, 50, and 100 nm. The narrow distributions demonstrate the periodicity. B) The number of monolayers per segment versus the diameter. Each data point represents an averaged value of 25-50 segments, taken from a single nanowire with $9.7*10^{-4}$ mbar DEZn doping. The curve is the theoretical expression given in the text, with $A = 0.70$ nm$^{-1}$ and $B = 6.5$ nm. C) The average segment length between two adjacent twin planes as a function of the Zn concentration. A transition from the Wz to ZB crystal structure is observed at $4.85*10^{-5}$ mbar DEZn, and above $4.85*10^{-4}$ mbar the twin planes have a constant spacing for a given dopant level. The data refers to wires with diameters between 15 and 25 nm. D) Calculated kinetic phase diagram showing the domains of Wz and ZB as a function of $\Delta\gamma/\gamma_{sl,ZB}$ and $\Delta\mu$. Zn reduces the liquid-solid step energy of ZB with respect to Wz and thus promotes the formation of zincblende InP wires.

**Figure 3: Model for periodic twinning in nanowires.** A) Schematic representation of the morphology of a twinned nanowire with the zincblende crystal structure with nonparallel {111} side facets. B) The cross-sectional shapes of the top facet of the nanowire crystal at the solid-liquid interface during growth. The numbers correspond to the positions indicated in A). Due to the non-parallel orientation of the side facets, {111}A edges increase and {111}B edges decrease in length during vertical growth, and as a result a hexagonal interface develops into a triangle-like shape. At a certain moment, it is energetically more favourable to create a twin plane rather than to continue growing towards a fully triangular top interface. After twin formation a triangle-like shape evolves to a hexagonal shape and the cycle is repeated as schematized in B). To the left or right the corresponding calculated shape of the catalyst particle on a hexagonal (1&3) and a triangularly deformed (2&4) interface is depicted, showing the skewing of the particle towards the long {111}A side edge and demonstrating that the contact angles depend sensitively on the cross-sectional shape.

**Figure 4: InP nanowire with alternating periodic and non-periodic segments.** A). An overview TEM image of a wire containing segments of intrinsic InP (containing randomly distributed stacking faults in a Wz structure) and Zn doped ($9.7*10^{-4}$ mbar) segments (with different lengths) with periodic twin planes in a ZB structure. B) Higher magnification TEM image of the segments closest to the gold particle. C) ZB section length versus the time interval during which the Zn precursor gas flow was switched on. The intercept at zero shows abrupt switching between the ZB and Wz crystal structure. D) Number of monolayers in each segment obtained from HRTEM. An average of ~ 13±3 monolayers is found.

# Figure 1

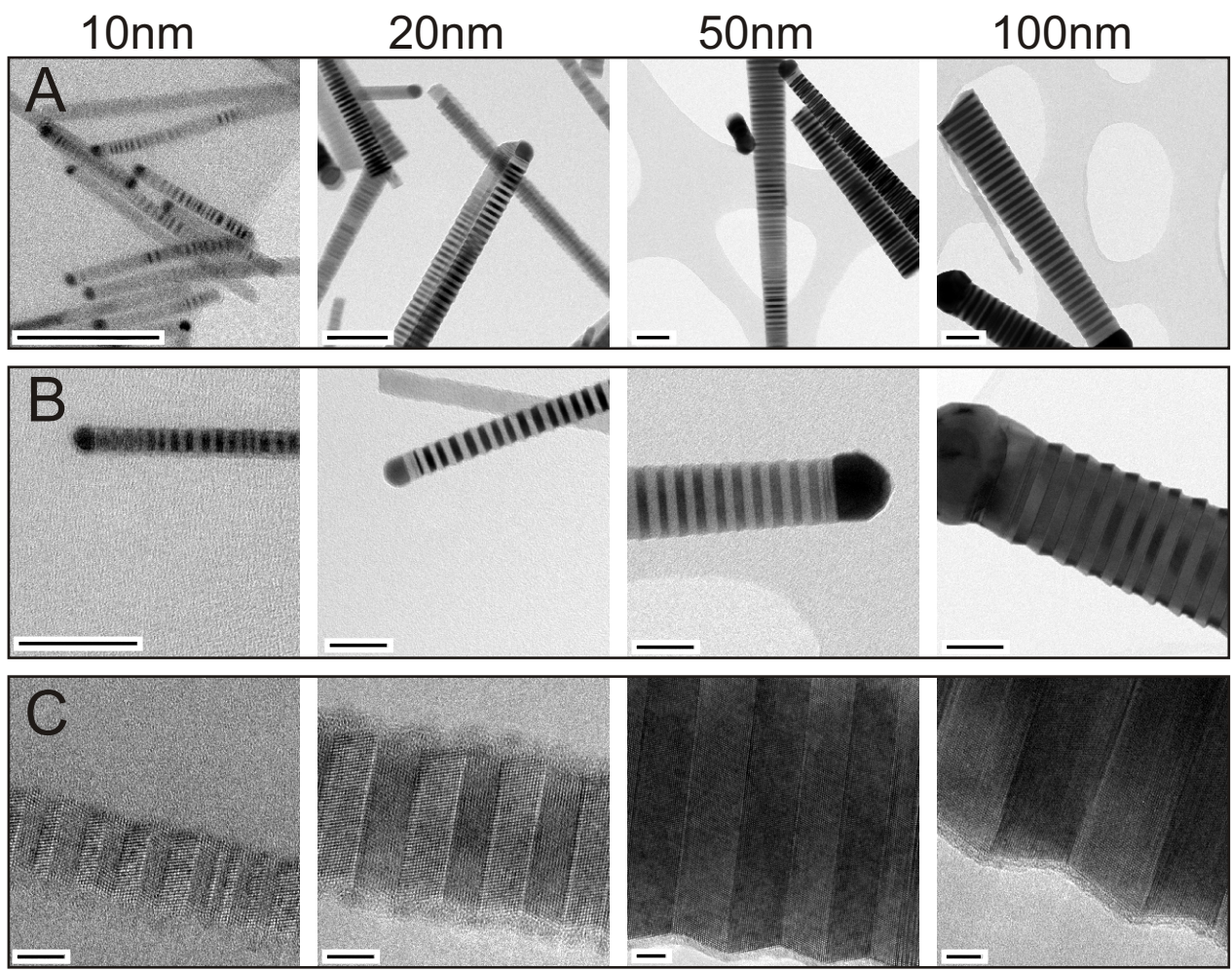

# Figure 2

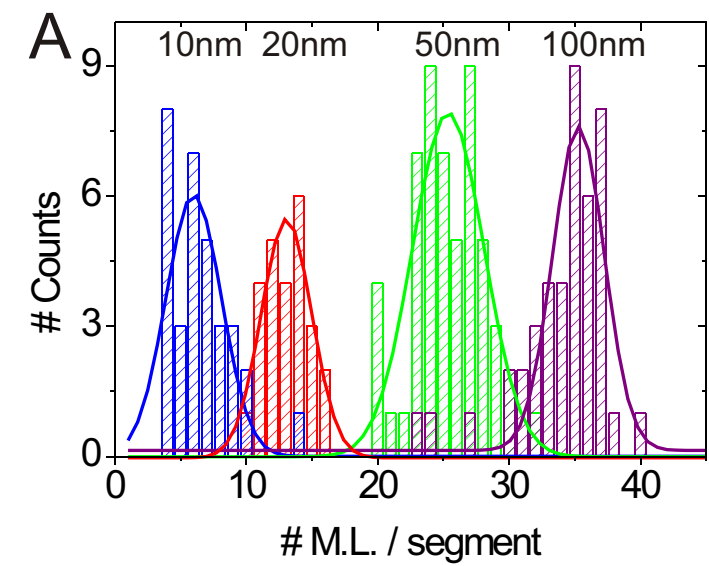
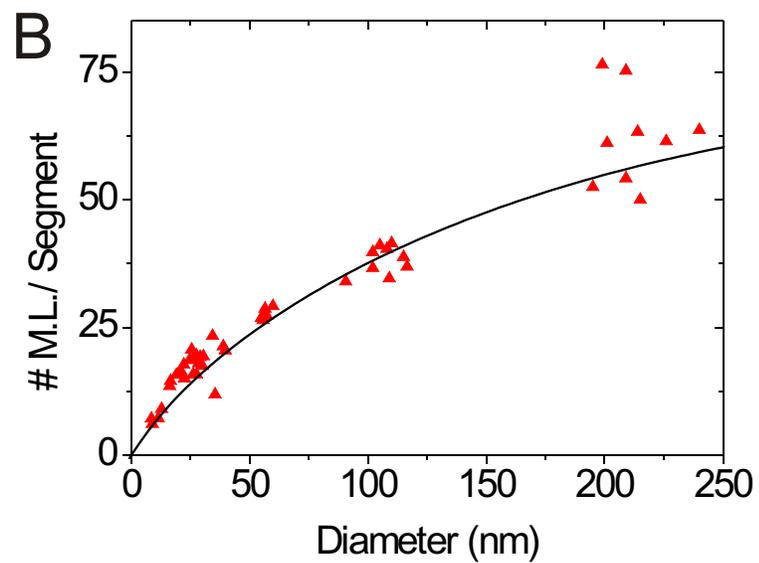
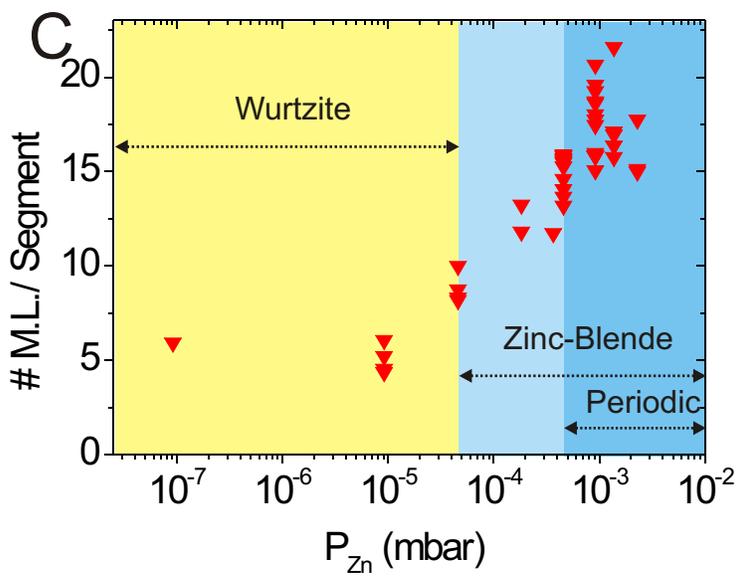
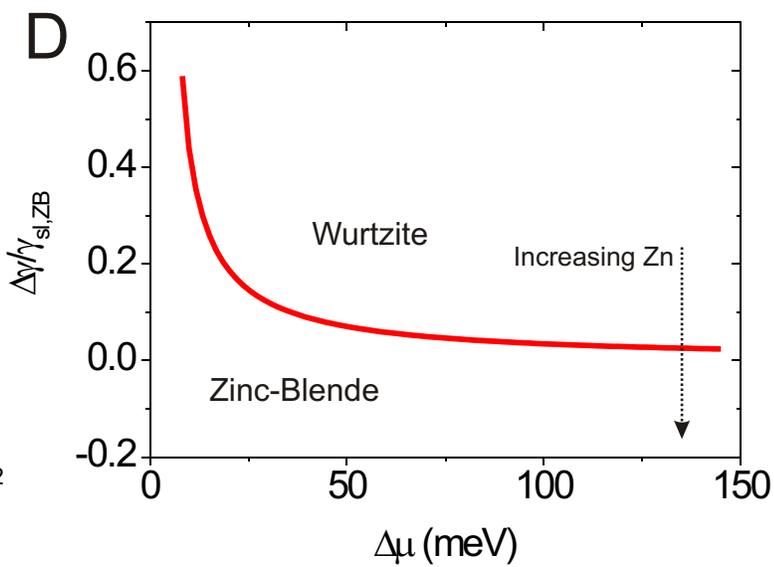

# Figure 3

A

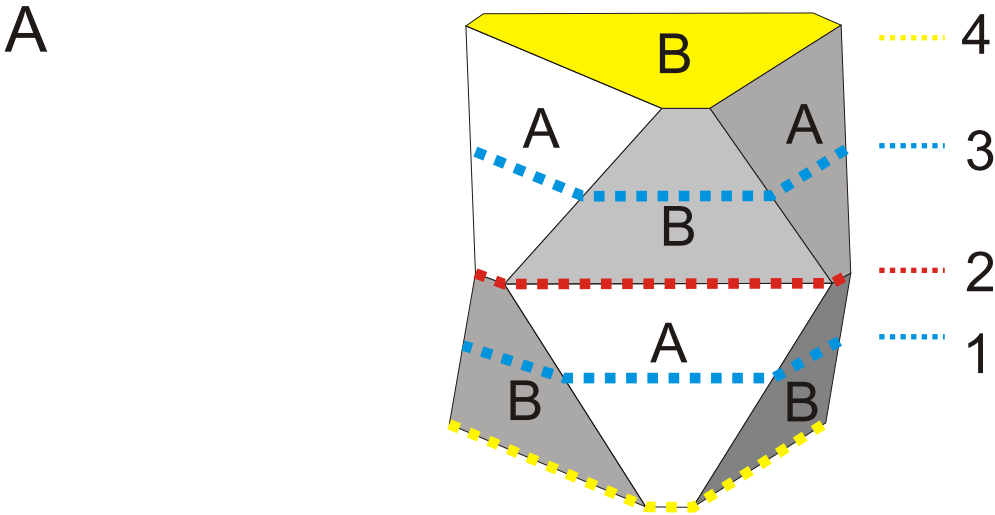

B

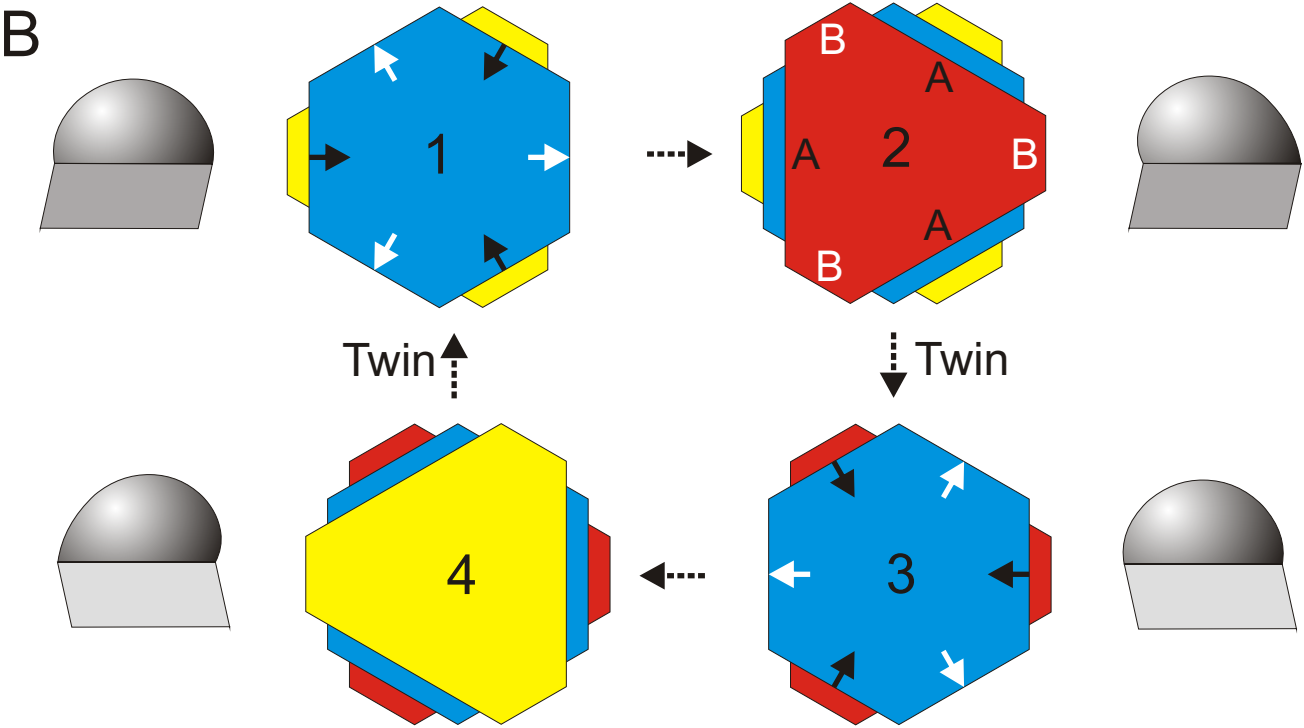

# Figure 4

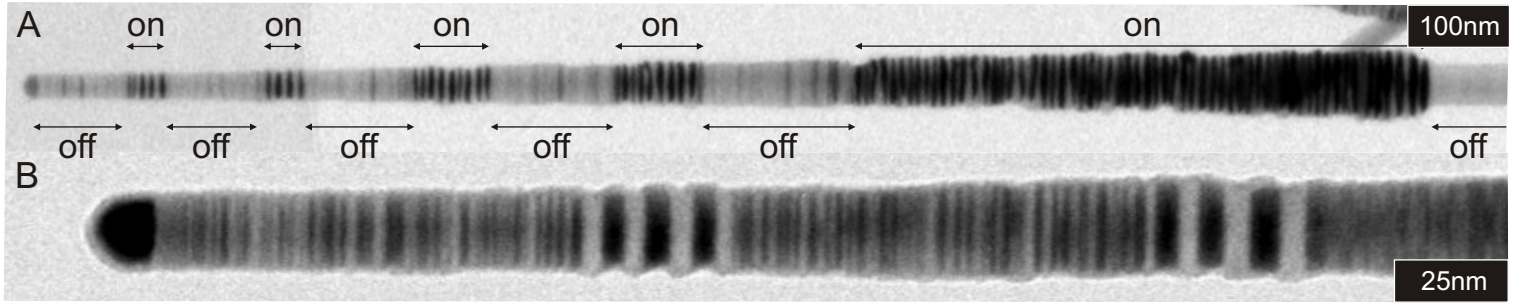

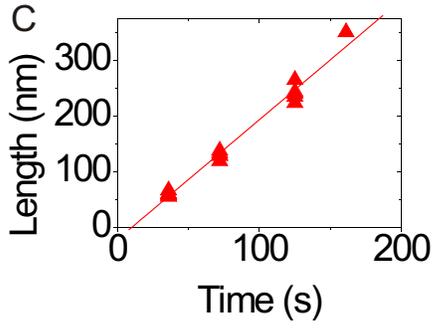
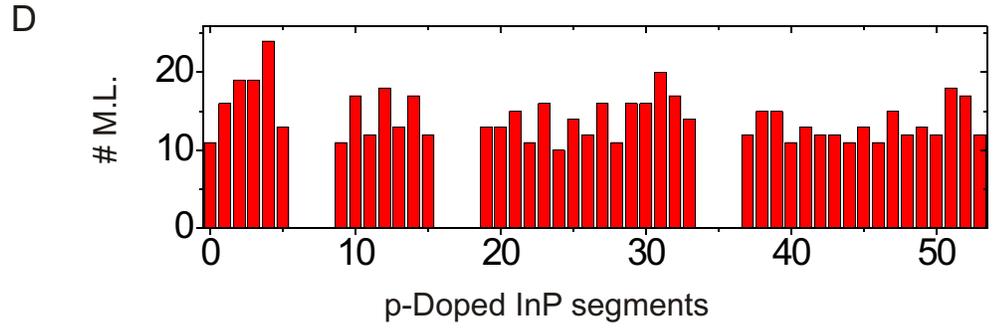